# Electronic band structure of narrow-bandgap p-n nanojunctions in heterostructured nanowires measured by electron energy loss spectroscopy

Reza R. Zamani,*,[1,2] Fredrik S. Hage,[3,4,5] Alberto Eljarrat,[6] Luna Namazi,[1] Quentin M. Ramasse,[3,7] and Kimberly A. Dick[1,8]

[1] *Solid-State Physics, Lund University, Box 118, Lund 22100, Sweden*

[2] *Department of Physics, Chalmers University of Technology, Gothenburg 41296, Sweden*

[3] *SuperSTEM Laboratory, SciTech Daresbury Campus, Daresbury WA4 4AD, United Kingdom*

[4] *Department of Materials, University of Oxford, Oxford OX1 3PH, United Kingdom*

[5] *Department of Physics / Centre for Materials Science and Nanotechnology, University of Oslo, Oslo 0318, Norway*

[6] *Institute of Physics, Humboldt University of Berlin, Berlin 12489, Germany*

[7] *School of Chemical and Process Engineering and School of Physics and Astronomy, University of Leeds, Leeds LS2 9JT, United Kingdom*

[8] *Centre for Analysis and Synthesis, Lund University, Box 124, Lund 22100, Sweden*

**Email:** reza.zamani@chalmers.se, reza.r.zamani@gmail.com

## Abstract

The electronic band structure of complex nanostructured semiconductors has a considerable effect on the final electronic and optical properties of the material and, ultimately, on the functionality of the devices incorporating them. Valence electron energy-loss spectroscopy (VEELS) in the transmission electron microscope (TEM) provides the possibility of measuring this property of semiconductors with high spatial resolution. However, it still represents a challenge for narrow-bandgap semiconductors, since an electron beam with low energy spread is required. Here we demonstrate that by means of monochromated VEELS we can study the electronic band structure of narrow-gap materials GaSb and InAs in the form of heterostructured nanowires, with bandgap values down to 0.5 eV, especially important for newly developed structures with unknown bandgaps. Using complex heterostructured InAs-GaSb nanowires, we determine a bandgap value of 0.54 eV for wurtzite InAs. Moreover, we directly compare the bandgaps of wurtzite and zinc-blende polytypes of GaSb in a single nanostructure, measured here as 0.84 and 0.75 eV, respectively. This allows us to solve an existing controversy in the band alignment between these structures arising from theoretical predictions. The findings demonstrate the potential of monochromated VEELS to provide a better understanding of the band alignment at the heterointerfaces of narrow-bandgap complex nanostructured materials with high spatial resolution. This is especially important for semiconductor device applications where even the slightest variations of the electronic band structure at the nanoscale can play a crucial role in their functionality.

**Main text**

The electronic band alignment at heterointerfaces in semiconductor nanomaterials is central to their performance in electronic and optical devices. It is particularly essential for complex arrangements of novel bandgap-engineered materials to be able to measure the bandgap locally and to determine the respective band alignments in different material configurations. The challenge is to successfully measure the bandgap value with high accuracy, and relevant spatial resolution.

Among the currently available techniques for measuring the bandgap of semiconductors, photoluminescence (PL) and optical absorption spectroscopy are the most common [1]. Although these methods provide high accuracy in energy measurement, they do not provide the necessary spatial resolution for observing the band structure in different parts of a complex heterostructure on the nanometer scale. VEELS, on the other hand, is a suitable technique with which we can measure the bandgap and reveal other optical properties of the material such as the complex refractive index, with very high spatial resolution down to a few atomic layers [2,3].

VEELS has been used to show the variation of the bandgap energy with the chemical composition changes across the interfaces of heterostructure layered semiconductor materials such as GaAsIn-GaAs [4] and grain boundaries in $Cu(In,Ga)Se_2$ [2]. Practical issues such as drift and mechanical stability have hampered the use of VEELS for bandgap measurements in nanowires and other low-dimensional objects. On the other hand, the size of the nanowires provides an advantage, with almost no risk of altering the material through sample preparation and with sample thickness intrinsically small, thus reducing plural scattering [5]. Moreover, other potential issues such as strong contribution from relativistic losses (Čerenkov radiation, guided light modes) are minimized [6,7].

Thanks to recent conceptual and technological advances in electron microscopy instrumentation, especially the introduction of highly stable monochromators enabling unprecedented energy spreads down to 5-8 meV [8,9], we are now able to measure very narrow bandgaps in semiconductors with high accuracy.

Here we exploit the capabilities of monochromated VEELS in order to study on the nanometer scale the electronic band structure of a complex heterostructure as schematically shown in Figure 1. The figure depicts the electron trajectory of the electron beam from the electron gun, passing through the monochromator and being energy-filtered by a narrow slit, and then interacting with the specimen, which in this case is a wurtzite InAs-GaSb core-shell nanowire with the shell covering half of the core and leaving a bare wurtzite InAs segment, then followed by an axial zinc-blende GaSb segment [10]. An important advantage of such a complex nanostructure is the availability of materials with known properties (on the same specimen with similar conditions, e.g. thickness, facets, crystal orientation) which can be used as reference to assure the accuracy of the measured unknown values.

***

Fig. 1a shows an overview HAADF-STEM image of the InAs-GaSb heterostructure nanowires which were grown by metal-organic vapor-phase epitaxy (MOVPE) as described in ref [10]. The lower segment of the nanowire is a partial radial heterostructured nanowire consisting of a wurtzite InAs core (red in Fig. 1b) partially covered with a tapered wurtzite GaSb shell (blue in Fig. 1b). Since the wurtzite GaSb shell (region i in the schematic in Fig.1b) does not cover the whole length of the wurtzite InAs core, there is a bare InAs nanowire (region ii) in the upper part of the lower segment. The wurtzite InAs is followed by a zinc-blende GaSb axial segment (region iii, green in Fig. 1b). The wurtzite polytype of GaSb is a metastable phase, which does not occur in bulk. In these nanowires, GaSb is forced to grow in the wurtzite structure via the so-called crystal phase transfer method [10]. Theoretically calculated bandgap values reported for this material are contradictory [11,12,13], and bandgaps both larger and smaller than the one of zinc-blende GaSb are predicted. Since the fabrication of this material was only very recently demonstrated, there have to date been no experimental reports of the electronic bandgap of wurtzite GaSb available to address this controversy. Here, the bandgaps of these different materials (wurtzite InAs, wurtzite GaSb and zinc-blende GaSb) can be measured directly in different regions of a single nanowire and correlated with atomic-resolution images of the structure obtained at the same time, thus, providing an unambiguous means of comparison.

In order to measure the bandgap and obtain information about the optical band structure, as mentioned, we use VEELS in a monochromated TEM. The schematic of Fig. 1c illustrates the trajectory of the electron beam in the device. The electrons, generated in a field-emission gun (bottom), travel through a set of prisms, become separated by their energy and then filtered by a slit. The monochromated electron beam then interacts with the specimen, and the exit electrons are collected by the spectrometer (top).

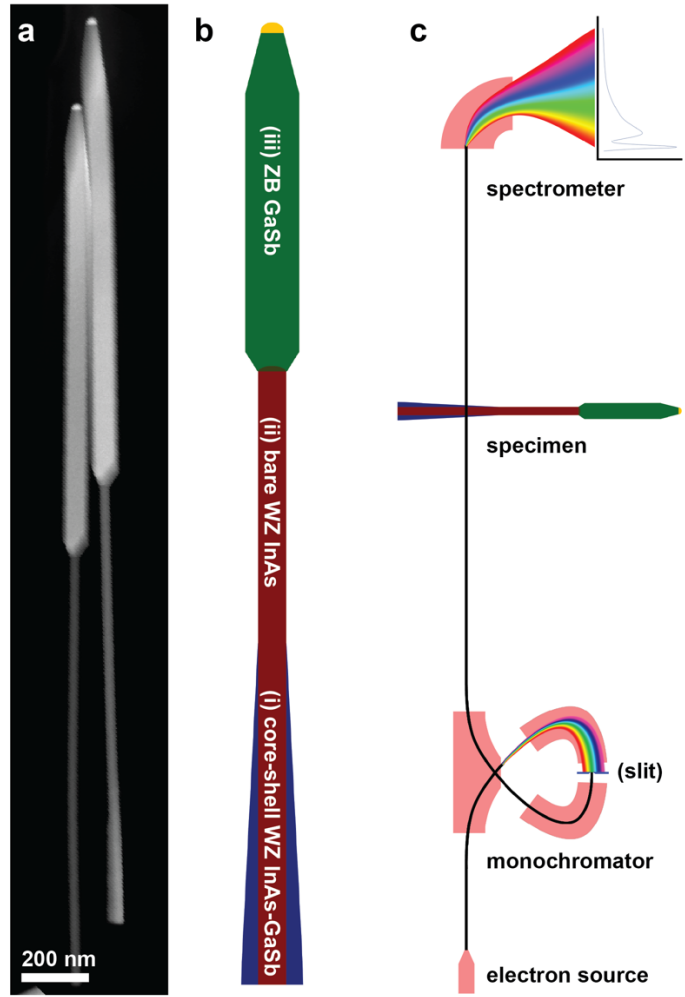

*Fig 1.* (a) HAADF-STEM image of the InAs-GaSb nanowires, (b) schematic of the material morphology, (c) schematics of the electron trajectory from the electron gun, to monochromator, then interacting with the specimen (InAs-GaSb nanowire) and being collected with the spectrometer

In the EEL spectrum of semiconductor or insulating materials, the bandgap edge appears as an intensity increase and change of curvature after the zero-loss peak (ZLP) and before higher energy-loss features such as plasmon peaks in the low-loss region, due to the absence of any permitted states that could be populated by electrons excited by the incoming beam. Therefore, the width of this region provides a measure of the bandgap of the material. Moreover, studying the energy loss function (ELF) directly after this region can provide an insight into the optical properties, related to the complex dielectric function (CDF) of the material [3]. Important experimental parameters, i.e. the convergence and collection angles of the beam, sample orientation and thickness, and the ZLP, must be considered however to interpret the value of the bandgap determined by VEELS. In particular, the decaying tails of the ZLP must be removed precisely in order to determine the onset of the first energy loss with high accuracy. Therefore, microscopes delivering an incident electron beam whose energy distribution has a narrow full-width half-maximum (FWHM, determining the energy resolution), and full-width tenth-maximum (FWTM, the tail of the ZLP which might overlap with the forbidden-transition region) are desired, especially for narrow-bandgap semiconductors. In a majority of TEMs, the tail of the ZLP extends to relatively high energy losses, sometimes covering the onset of the first plasmon peaks and thus resulting in relatively large errors in bandgap measurements [6]. Beam monochromators narrow down the energy spread, as shown schematically in Fig. 1c, which results in a considerably narrower ZLP (down to <10 meV [8]) with a smaller tail, and provides much higher reliability in bandgap measurement, especially for narrow-gap materials.

Here, rather than closing the monochromating slit to its narrowest setting for ultimate energy resolution, we chose a wider slit size for a ZLP FWHM of 30-40 meV, sufficient for bandgap measurements while providing beam currents high enough for high signal-to-noise bandgap measurements. Further experimental details can be found in the Methods section.

Determination of the bandgap of the bare wurtzite InAs section of the heterostructured nanowire (Fig 2a) is shown by the low-loss EEL spectrum shown in Fig 2. The thickness of this part of the nanowire in projection is around 40 nm as shown in the HAADF STEM image of Fig 2b. The atomic resolution HAADF STEM image of Fig. 2c reveals that the nanowire is oriented in the [1-100] direction. The blue line in Fig. 2d is the raw experimental EEL spectrum, in the sense that neither subtraction of the ZLP nor smoothing have been applied. A model is fitted to the raw spectrum, that contains a power-law component for the ZLP tail and a square-root function for the direct bandgap contribution (further details about the fit procedure are provided in the Methods section). Because InAs is known to possess a direct bandgap in both wurtzite and zinc-blende structures, the square-root dependency for the EELS intensity near the intensity onset is expected, whose intercept with the energy axis is taken as a measure of the bandgap value [14]. The fitted curve, shown in Fig. 2d as a blue dashed line, indicates the onset at 0.54 eV agreeing reasonably well with the expected value for the bandgap of wurtzite InAs [15]. This value can be used as an internal reference for the parts of the material with unknown band structures. In fact, an exact match with values obtained by alternative experimental techniques is not expected, as instrumental parameters will influence the fine details of the recorded EEL spectrum. Nonetheless, in otherwise identical conditions, relative differences in the determined gap onset can be used to quantify bandgap differences between different materials (or different parts of a same sample, at grain boundaries, for instance [4,16]).

The energy losses in the region immediately after the intensity onset indicating the bandgap (0.5-8.0 eV), as shown in Fig 2e, correspond well to the transitions expected from the complex dielectric function of the wurtzite InAs in the in-plane orientation [17]. This is also confirmed by the good agreement with relativistic EELS calculations obtained using the Kröger formula [3] while taking into account the appropriate experimental conditions.

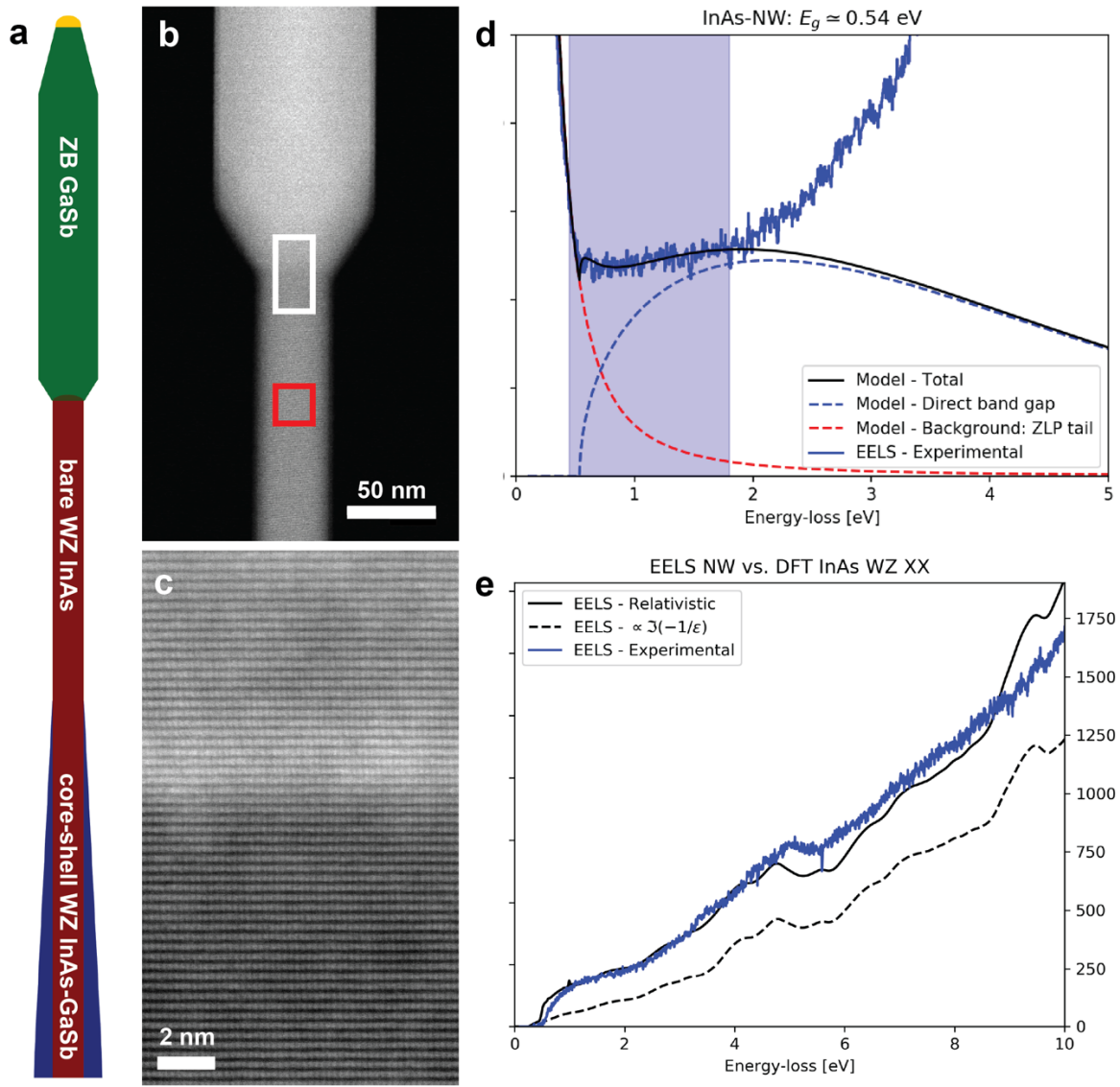

*Fig 2.* VEELS on wurtzite InAs: (a) Schematic of the InAs-GaSb nanowires, (b) Overview HAADF-STEM image depicting the bare wurtzite InAs segment (below) and its axial interface with the zinc-blende GaSb segment (above), (c) atomic resolution HAADF-STEM image of the same heterointerface from the [1-100] zone axis, (d) experimental low-loss EEL spectrum of wurtzite InAs nanowire (blue line) together with the model for ZLP subtraction (red dash line), model for direct bandgap (blue dash line) and the total (black line), € experimental low-loss EEL spectrum (blue line) together with the non-relativistic (black dash line) and relativistic (black line) ELF models

Fig. 3 illustrates a similar analysis for spectra acquired in the nanowire sections corresponding to the zinc-blende and metastable wurtzite phases of GaSb indicated on the schematic of the heterostructures nanowire in Fig. 3a. The top part in green depicts the zinc-blende GaSb axial segment on the InAs nanowire (HAADF STEM image in Fig. 3b), and the lower part of the lower segment shows the wurtzite GaSb shell around the InAs core (HAADF STEM image in Fig 3e). Fig. 3c shows the low-loss EEL spectrum on the zinc-blende GaSb segment. As in the case of InAs shown above, the blue line corresponds to the experimental spectrum, the red dashed line to the ZLP tail component of the fit model and the blue dashed line to the square-root function whose onset provides a measure of the direct bandgap of zinc-blende GaSb. The green dashed line is a further component of the fit model, which addresses the pre-onset intensity using an indirect transition model with a sigmoid-like decay. As can be seen, the onset of the best fit parabola indicates a bandgap value of 0.75 eV, in good agreement with the previously reported value in GaSb (0.725 eV in bulk at room temperature [18,19]). On the other hand, Fig. 3f reveals the bandgap value of wurtzite GaSb, to be at 0.84 eV. This value was measured in the lower parts of the nanowire, through a relatively thick section of the shell (up to 30-40 nm thick, comparable to the projected thickness of the zinc-blende GaSb section studied above); the contribution of the InAs core signal is assumed to be minimized by placing the beam away from the InAs region. This direct comparison in a controlled geometry appears to solve the discrepancy in reported bandgap values for wurtzite GaSb in earlier theoretical studies, with values as narrow as 0.503 eV [11,20] or as wide as 0.835 eV [12]. Although the absolute value determined in our experiments may not be directly comparable to other experimental measurements (e.g. with PL or ellipsometry), by probing directly in identical experimental conditions the two polytypes we demonstrate conclusively that the bandgap of wurtzite GaSb is around 0.1 eV wider than its zinc-blende counterpart, consistent with the prediction by Belabes et al. [12].

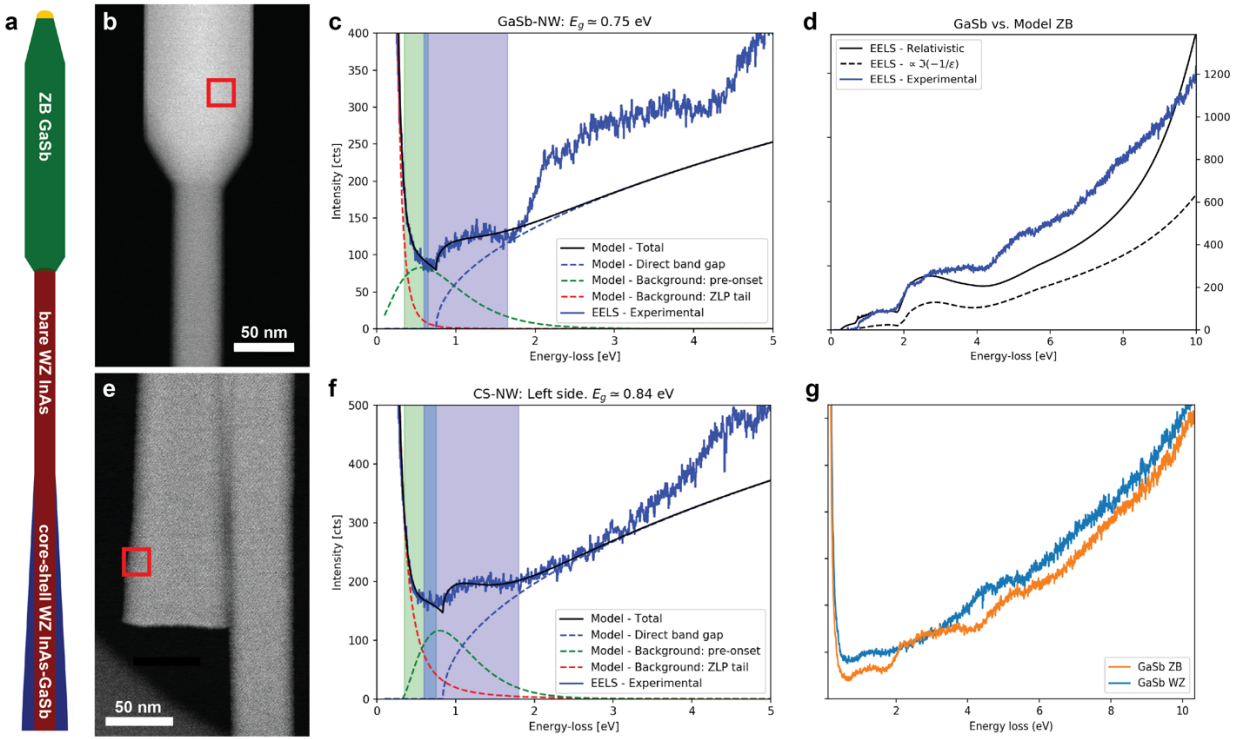

*Fig 3.* VEELS on wurtzite and zinc-blende GaSb: (a) Schematic of the InAs-GaSb nanowires, (b) Overview HAADF-STEM image depicting the bare wurtzite InAs segment (below) and its axial interface with the zinc-blende GaSb segment (above), (c) experimental low-loss EEL spectrum of zinc-blende GaSb segment (blue line) together with the model for ZLP subtraction (red dashed line), model for direct bandgap (blue dashed line), pre-onset intensity model for indirect transition (green dashed line) and the total (black line), (d) experimental low-loss EEL spectrum (blue line) together with the non-relativistic (black dashed line) and relativistic (black line) ELF models, ∈ Overview HAADF-STEM image depicting the core-shell wurtzite InAs-GaSb nanowire (left), (f) experimental low-loss EEL spectrum of wurtzite GaSb shell (blue line) together with the model for ZLP subtraction (red dashed line), model for direct bandgap (blue dashed line), pre-onset intensity model for indirect transition (green dashed line) and the total (black line), (g) comparison of the experimental EEL spectra of wurtzite and zinc-blende GaSb.

Apart from the different bandgap energy, there are differences observed in other peaks in the spectrum. In particular the strong direct transition right after the bandgap-indicating onset, observed below 2 eV in zinc-blende GaSb (see Fig. 3d), is also observed in the wurtzite GaSb. However, it appears to be weaker and seem to start at around 3 eV. This discrepancy could be due to the different band structures of wurtzite and zinc-blende GaSb, which would require further investigation.

***

In summary, we demonstrated the application of monochromated VEELS for measuring the electronic band structure of InAs-GaSb heterostructured nanowires at the nanometer scale. The ideal geometry of these nanowires allows us to determine the unknown bandgap of wurtzite GaSb, readily comparable to the one of its zinc-blende counterpart (0.1 eV wider for wurtzite), and therefore, avoid discrepancies and controversies. This method can be used for studying the electronic band structure of a wide range of semiconductors with the energy and spatial resolutions required for complex configurations where the changes at the atomic level can alter the properties of the material in macro scale. Monochromated VEELS in combination with other electron microscopy methods such as core-loss EELS mapping and aberration-corrected STEM imaging, can provide crucial information about the physical properties of materials and predict their behavior in novel semiconductor devices such as quantum computers.

**Acknowledgements**

The authors acknowledge the financial support by the European Research Council under the European Union's Seventh Framework Programme (FP/2007-2013)/ERC grant agreement no. 336126; the Swedish Research Council (VR); the Knut and Alice Wallenberg Foundation (KAW); SuperSTEM is the UK National Research Facility for Advanced Electron Microscopy, supported by the Engineering and Physical Sciences Research Council (EPSRC). A.E. acknowledges funding by the Deutsche Forschungsgemeinschaft (DFG) – Project no. 182087777 - SFB951. Luis C. O. Dacal is acknowledged for constructive discussions.

**Methods**

**Nanowire growth –** The nanowires are grown by metal-organic vapor phase epitaxy (MOVPE), using aerosol Au nanoparticles as catalysts on InAs (111)B substrates. First, wurtzite InAs nanowires are grown. Afterward, GaSb is deposited. GaSb forms a zinc-blende axial segment, as well as a tapered wurtzite shell on the wurtzite InAs cores (following the same crystal structure). The full growth procedure is described in a previous report [1].

**TEM specimen preparation –** The nanowires are transferred from the substrate to lacey carbon grids by mechanical approximation. Thereafter, they the specimen is baked for several hours at an elevated temperature (~130°C) in high vacuum (~2x10$^{-6}$ Torr) prior to the insertion to the electron microscope, in order to gently remove hydrocarbon contaminations without damaging the nanowires.

**EELS –** The beam is monochromated in a way that energy resolutions of 30-40 meV are achieved, while an energy dispersion of 5meV/channel was used for the spectra presented in the main text. It is possible to reach much higher energy resolutions (down to 5-8 meV [2,]) by decreasing the monochromator's slit width, however, this is not ideal for our experiments as the electron beam intensity, and consequently signal to noise ratio, decrease dramatically. The acceleration voltage was 60 kV. The probe convergence and EEL spectrometer aperture semi-angle were 31 mrad and 44 mrad, respectively. While including the intense ZLP in a recorded spectrum aids in calibration and estimation of the energy resolution, the finite dynamic range of the CCD EELS camera used was found to severely limits the signal-to-noise of the comparatively much weaker signal in the bad-gap region. To address this, we used the so-called "Dual-EELS" approach [3], where two spectra including different energy loss ranges are acquired in rapid succession, for each pixel in an EEL spectrum image. Acquisition times can be independently chosen for these two spectra so that the signal-to-noise is optimized for each. In our case we chose to acquire spectra of the band gap region where (1) the full ZLP was included and where (2) the ZLP maximum was shifted off the camera so that only parts of the energy loss tail of the ZLP was detected.

Line scans are acquired on target regions of the nanowires. It is important to avoid the amorphous carbon film of the TEM grid as this will significantly complicate the interpretation of resulting spectra. Therefore, the spectra are acquired on nanowires in which the target regions are hanging on vacuum, several tens of nanometers away from the amorphous carbon film.

**ZLP subtraction –** An important parameter to take into account is the consistency of the shape of the ZLP as it is required to be subtracted. In order to ensure that the ZLP is subtracted correctly, it is recorded in identical conditions (within the same dataset) in vacuum, at least 100 nm away from the nanowire and the carbon film. This is used as a reference to model and subtract the ZLP from the low-loss spectrum obtained on the nanowire after scaling. The resulting spectra are compared to the relativistic EELS calculations. The methodology described by Rafferty and Brown is used in order to determine the band gap value [4]. This procedure is applied in Figures 2e, and 3d,g.

**Spectrum fitting –** Since InAs and GaSb materials have direct bandgaps in both wurtzite and zinc-blende structures, a square root model, $(E-E_g)^{1/2}$, is used to fit the spectrum near the intensity onset that reveals the bandgap value [4]. Apart from the square-root function, the shape of the EELS spectra near the intensity onset contains several other contributions. First and foremost, the contribution of the ZLP tail has to be considered. We use a power-law background that contains only a few parameters but provides enough flexibility to extrapolate this contribution from a short range before the onset indicating the bandgap. Additionally, we observe some pre-onset intensity before the direct bandgap edge in GaSb in the experimental data, not reproduced in the relativistic calculations, and thus likely unrelated to retardation losses. The origin of this intensity is not clear, however, we propose that it is mainly due to indirect transitions before the direct bandgap onset from defects in the crystalline structure or few ångströms of amorphous material around the nanowire. For this reason, we model it using an indirect transition model that follows a $(E-E_g)^{3/2}$ dependency [4]. Finally, the EELS intensity away from these transitions is allowed to decay using Sigmoid-like functions (Fermi-Dirac distribution), improving the agreement with the experimental spectra. These contributions are fitted in a sequential fashion. First, the power-law parameters estimated from the pre-onset region. Direct and indirect band gap are pre-fitted using the Levenberg-Marquardt algorithm. Finally, dual annealing is employed to optimize the parameters of the final model (5% allowed variation) [5]. HyperSpy is used for coding [6].

The standard deviations for the measured direct bandgaps are usually in the order of the energy dispersion which was around 5 meV/ch. Nevertheless, as the ZLP width was larger than that and thus energy resolution is ultimately around 30 meV. Therefore, for the bandgap values with a standard deviation within the limits of our instrumental resolution, the tolerance can be ±3 meV.

**Methods References**